\documentclass[12pt]{article}

\usepackage{simplemargins,hyperref,authblk,graphicx,amssymb,amsmath,multirow,booktabs,xcolor,xr,lineno,pdflscape}

\usepackage[super,comma]{natbib}

\usepackage{etoolbox}

\setallmargins{1in}

\date{}

\title{Episodic deluges in simulated hothouse climates}
\author[1]{\small Jacob T. Seeley}
\author[1,2]{\small Robin D. Wordsworth}

\affil[1]{\footnotesize Department of Earth and Planetary Sciences, Harvard University, Cambridge, MA}
\affil[2]{\footnotesize School of Engineering and Applied Sciences, Harvard University, Cambridge, MA}

\begin{document}
\maketitle

\textbf{Earth's distant past and potentially its future include extremely warm ``hothouse''\citep{Steffen2018} climate states, but little is known about how the atmosphere behaves in such states. One distinguishing characteristic of hothouse climates is that they feature lower-tropospheric radiative heating, rather than cooling, due to the closing of the water vapor infrared window regions\citep{Wordsworth2013}. Previous work has suggested that this could lead to temperature inversions and significant changes in cloud cover\citep{Wolf2015,Kopparapu2016,Popp2016,Wolf2018}, but no previous modeling of the hothouse regime has resolved convective-scale turbulent air motions and cloud cover directly, thus leaving many questions about hothouse radiative heating unanswered. Here, we conduct simulations that explicitly resolve convection and find that lower-tropospheric radiative heating in hothouse climates causes the hydrologic cycle to shift from a quasi-steady regime to a ``relaxation oscillator'' regime, in which precipitation occurs in short and intense outbursts separated by multi-day dry spells. The transition to the oscillatory regime is accompanied by strongly enhanced local precipitation fluxes, a significant increase in cloud cover, and a transiently positive (unstable) climate feedback parameter. Our results indicate that hothouse climates may feature a novel form of ``temporal" convective self-organization, with implications for both cloud coverage and erosion processes.}

For the past few million years, Earth’s climate has been characterized by fairly cool conditions, with repeated transitions between glacial and interglacial climates\citep{Snyder2016}. On longer timescales, however, the range of Earth's climate states is far wider. In the Hadean and Archean\citep{Sleep2010,Charnay2017}, as well as in the aftermath of Neoproterozoic Snowball events\citep{Pierrehumbert2011}, high carbon dioxide levels may have elevated surface temperatures by tens of degrees Kelvin compared to today. In the distant future, increases in solar luminosity will cause surface temperature to increase and eventually drive Earth through a runaway greenhouse transition\citep{Goldblatt2012}. No matter the forcing mechanism, warming of Earth's climate causes atmospheric water vapor to accumulate 
rapidly, strengthening the water vapor greenhouse effect by rendering more of the infrared spectrum opaque. With sufficient warming, even the most weakly absorbing spectral ``window'' regions in the thermal infrared are closed off and the lower troposphere can no longer cool by emitting infrared radiation to space\citep{Koll2018,Seeley2021}. However, because water vapor also absorbs in the near-infrared \citep{Wordsworth2013}, tropospheric absorption of incoming solar radiation persists in extremely warm climates, eventually yielding net lower-tropospheric radiative heating (LTRH). This phenomenon may also occur on the intensely-irradiated day sides of tidally-locked exoplanets\citep{Kopparapu2016}, the climates of which are strongly influenced by convection near the substellar point \citep{Yang2013a,Sergeev2020,Lefevre2021}.

\section*{Simulations of hothouse climates}

We investigated hothouse climates using a convection-resolving model in which radiative and convective heating rates are constrained to be in time-mean balance \cite{methods}. The model we employ (DAM\cite{romps2008}) is fully compressible and nonhydrostatic, and is well suited to simulating very warm climates in two key respects: 1) it takes into account the full thermodynamics of moist air, including changes in atmospheric pressure due to condensation and the effect of water on the heat capacity of air; and 2) the radiative transfer scheme has been modified to remain accurate in very warm climates\citep{methods}. For simulations with a time-evolving sea surface temperature (SST), the model SST is evolved according to the sum of surface enthalpy fluxes, radiative fluxes, and a prescribed ocean heat sink; we also performed simulations with fixed SST. Our baseline simulations were conducted on a 72 km $\times$ 72 km square grid with doubly-periodic horizontal boundary conditions. Further information about our model setup is available in the Methods section\citep{methods}.

\begin{figure*}[ht!]
\centerline{\includegraphics[width=\textwidth]{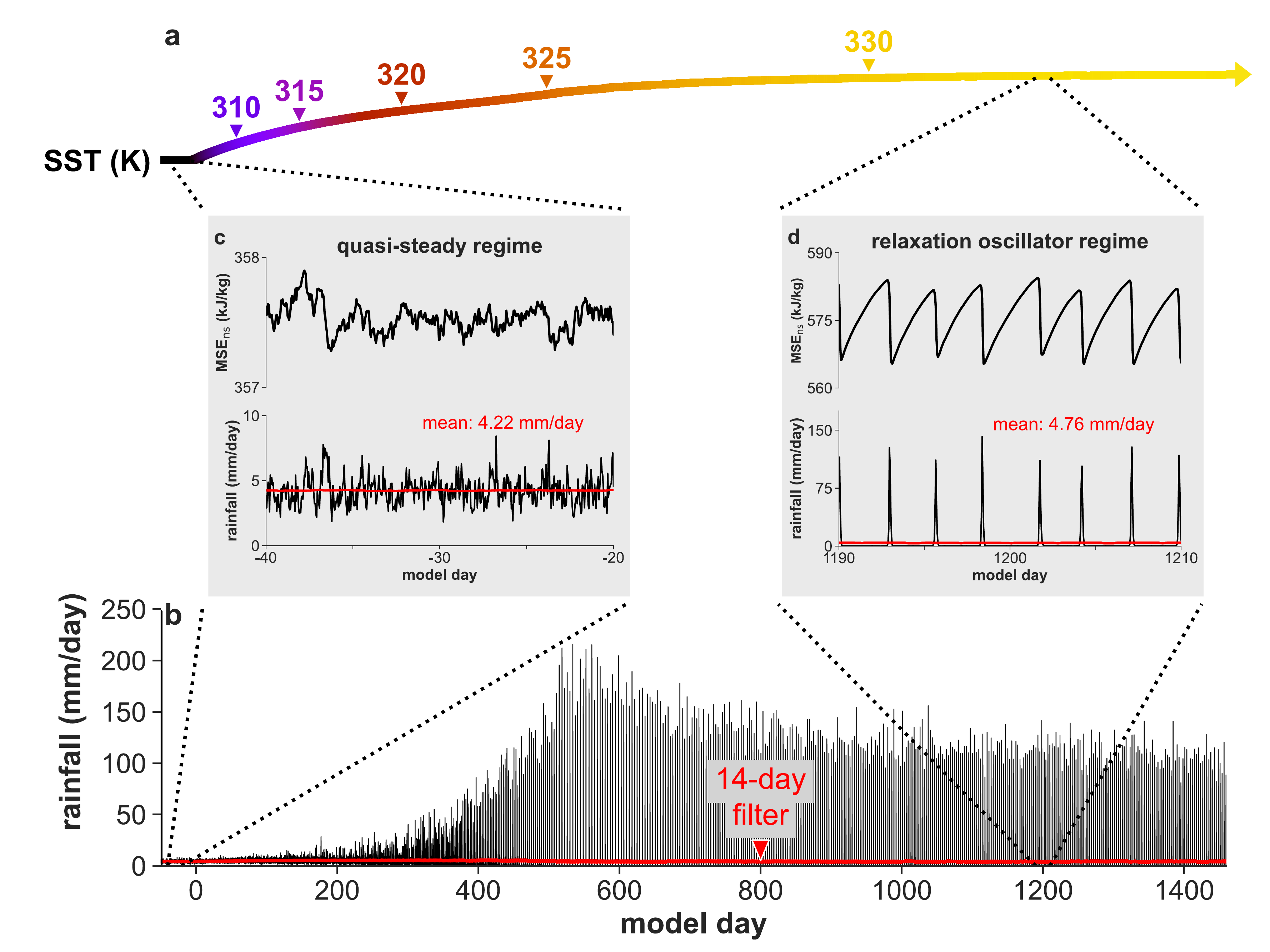}}
\caption{\textbf{Transition to the relaxation oscillator regime due to increased insolation.} Results are from a simulation initiated from an equilibrated state with a mean SST of 305 K but with a 10\% increase in the solar constant, imposed on model day 0 in this figure. (a) 14-day-filtered SST with first-crossings indicated for 5-K increments above 305 K. (b) Hourly (black) and 14-day-filtered (red) precipitation. (c-d) Hourly precipitation and mean near-surface ($z<1$ km) moist static energy (MSE$_\mathrm{ns}$) over 20-day intervals before and after the transition to the relaxation oscillator regime. Moist static energy is defined as $\mathrm{MSE}=c_p T + L q_v + g z$, where $c_p$ (J/kg/K) is the heat capacity of air at constant pressure, $T$ (K) is the temperature, $L$ (J/kg) is the latent heat of vaporization, $q_\mathrm{v}$ (kg/kg) is the specific humidity, $g$ (m/s) is the gravitational acceleration, and $z$ (m) is the altitude.}
\label{fig:sst_precip}
\end{figure*}

Figure 1 shows results from a convection-resolving simulation that was initiated with an SST comparable to the warmest on Earth today (305 K, or 32 $^\circ$C), but with a 10\% increase in the solar constant --- equivalent to Earth about 1 billion years in the future, or on an orbit about 5\% closer to the Sun today. In response to this large solar forcing, the model warms rapidly and reaches a new equilibrium at a mean SST of about 330 K (57 $^\circ$C) within four years of model-time. A fundamental shift in state is evident in Figures 1b--d, which show time series of hourly precipitation and near-surface moist static energy during the model evolution. The early stages of the simulation are in a quasi-steady convective regime, with hourly precipitation rates exhibiting noisy fluctuations about a mean of approximately 5 mm/day. Later in the simulation, as the SST warms above about 320 K, the nature of the precipitation changes fundamentally. Rather than exhibiting noisy fluctuations about the mean, precipitation occurs in large outbursts lasting a few hours, separated by regular multi-day dry spells during which there is essentially no precipitation. As in a canonical relaxation oscillator\citep{Wang1999,Ginoux2012}, the latter regime is characterized by repeated sequences of slow destabilization and fast stabilization (Fig. 1d); hence we refer to this hothouse state as a ``relaxation oscillator'' or ``oscillatory" regime. Since our simulations receive constant diurnally-averaged insolation\citep{methods}, the emergence of periodic behavior must be the result of radiative-convective feedbacks. Across the transition to the relaxation oscillator regime, the intensifying dry spells balance the increasingly large outbursts of precipitation such that time-mean rainfall before and after the transition changes by only about 10\% (Figure 1b--d, 14-day-filtered rainfall), as predicted by mean energetic constraints\citep{pendergrass2014,jeevanjee2018b}. However, the maximum domain-mean hourly rain rates increase by an order of magnitude in the oscillatory regime. 

While Figure 1 shows that the oscillatory regime can result from solar-forced warming, the closing of the water vapor spectral windows is fundamentally tied to increased temperature and should occur regardless of the forcing mechanism. Indeed, our tests suggest that the oscillatory state is a general property of very warm and moist atmospheres simulated by convection-resolving models, being robust to forcing mechanism, microphysics scheme, domain size and resolution, and choice of convection-resolving model (Fig. E3). To further confirm the central importance of lower-tropospheric radiative heating, we conducted simulations across a range of surface temperatures with idealized time-invariant radiative heating profiles that resemble either cool-climate conditions (with radiative cooling throughout the troposphere) or hothouse conditions (with lower-tropospheric radiative heating). The cool-climate radiative heating profiles produce quasi-steady convective behavior at all temperatures (LTRH$\_$off; Fig. 2a,d), whereas the hothouse-type radiative heating profiles produce the relaxation oscillator regime at all temperatures (LTRH$\_$on; Fig. 2c,f). The simulations with radiation calculated interactively interpolate between these two regimes (Fig. 2b,e), suggesting that lower-tropospheric radiative heating is the key characteristic of hothouse climates that drives the transition to the relaxation oscillator regime.

\begin{figure*}[ht!]
\centerline{\includegraphics[width=0.9\textwidth]{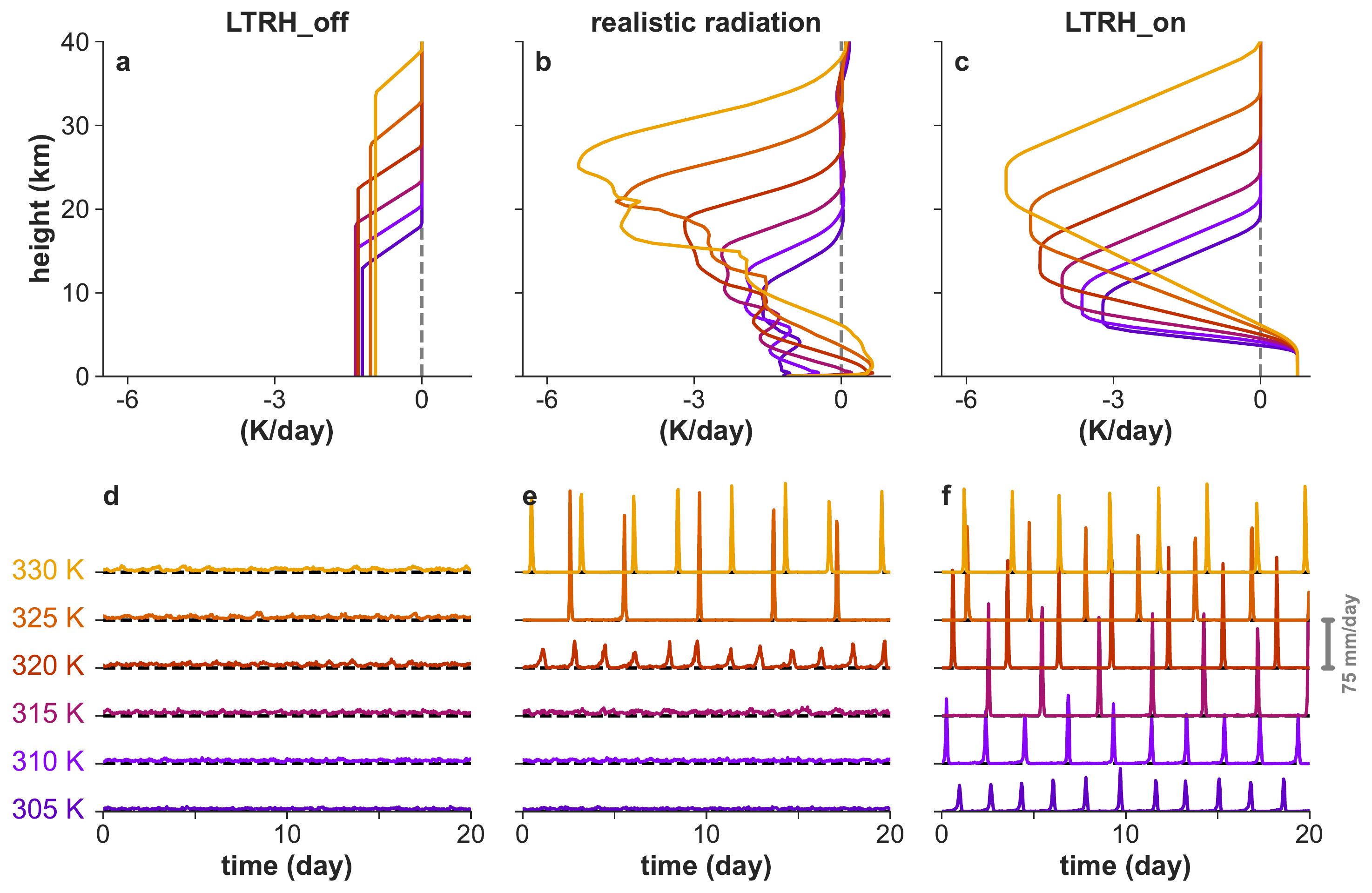}}
\caption{\textbf{The oscillatory regime is induced by lower-tropospheric radiative heating.} (Top row) Vertical profiles of radiative heating from the (a) LTRH\_off, (b) fixedSST, and (c) LTRH\_on simulations\citep{methods}. The fixedSST simulations use realistic radiative transfer calculated with the model-generated vertical profiles of temperature and absorber densities, while for the other two experiments, the radiative heating profiles are prescribed to resemble either cool-climate conditions (radiative cooling throughout the troposphere; LTRH\_off) or hothouse conditions (radiative heating in the lower troposphere; LTRH\_on). Cool to warm colors indicate increasing SST. (Bottom row, panels d--e) 20-day time series of domain-mean precipitation from the simulations. For visual clarity, the precipitation data is offset vertically by 75 mm/day (gray scale bar at right) for each 5-K increment of SST.}
\label{fig:LTRH}
\end{figure*}

\section*{Physical basis of the oscillatory regime}

To reveal the mechanism behind the oscillatory state, we studied high-frequency output from fixed-SST simulations at 330 K. This output shows that the oscillatory state consists of three main phases: recharge, triggering, and discharge (Fig. 3). An animation of model output in the oscillatory state can be viewed in the Supplementary Video, and a summary schematic of the phases of the oscillatory state is presented in Figure 4.

\begin{figure*}[ht!]
\centerline{\includegraphics[width=\textwidth]{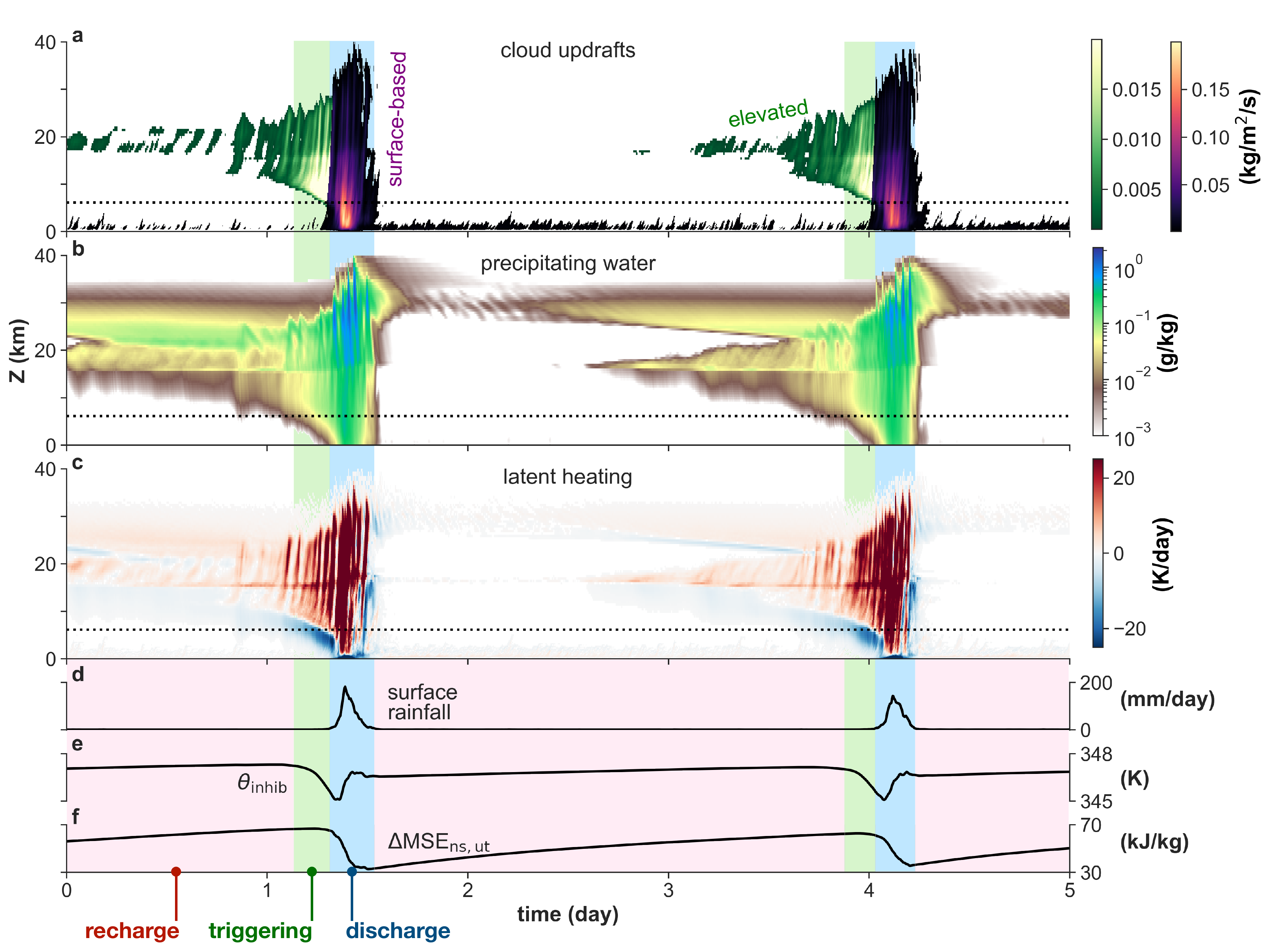}}
\caption{\textbf{Mechanism of the oscillatory regime as revealed by high-frequency model output.} (a--c) Time-vs-height plots of (a) cloud updraft mass flux, (b) precipitating water mass fraction $q_p$, and (c) latent heating from the fixedSST simulation at 330 K. Convective mass flux was divided into ``surface-based" and ``elevated" categories using a passive tracer\citep{methods}. (d--f) Timeseries of (d) domain-mean surface precipitation, (e) $\theta_\mathrm{inhib}$, the mean potential temperature in the inhibition layer ($2000 < z < 5500$ m), and (f) difference in mean moist static energy between the near-surface layer ($z<1$ km) and the upper troposphere ($25 < z < 35$ km) from the same simulation. The discharge phases in this figure (blue shading) are identified as intervals with a surface precipitation rate above 5 mm/day. The triggering phases (green shading) begin when the hourly-mean latent heating rate in the inhibition layer is more negative than -2 K/day and the surface precipitation rate is less than 1 mm/day, and end when the ensuing discharge period commences. The recharge phases occur between the end of a discharge phase and the beginning of a triggering phase. The dotted horizontal lines in panels a--c marks the level of zero radiative heating (i.e., the transition from time-mean radiative heating to cooling).}
\label{fig:hifreq}
\end{figure*}

During an outburst of precipitation (discharge phase), the lower troposphere is flooded with negatively-buoyant downdrafts of cold and dry air with low moist static energy (Fig. 3f, Supplementary Video). Over the course of the ensuing recharge phase, LTRH keeps the surface and the upper troposphere decoupled by increasing the mean potential temperature in the intervening ``inhibition layer'' (Fig. 3e) and suppressing surface buoyancy fluxes. With the inhibition layer effectively throttling surface-based convection, surface evaporation humidifies the near-surface air without any compensating ventilation into the upper troposphere (Fig. 3a), while radiative cooling aloft cools the upper troposphere; combined, these processes lead to a large build-up of convective instability (Fig. 3f). Over the course of a few days, this build-up leaves the atmosphere primed for an intense precipitation event --- a powder keg ready to explode.

The explosion of the powder keg is ultimately triggered from the top down by the influence of elevated convection. In the absence of subsidence warming and drying that would keep clear air unsaturated, radiative cooling aloft during the recharge phase leads to in-situ cooling, condensation, and elevated convection with cloud bases above 7 km. This elevated convection produces virga (precipitation that evaporates before reaching the ground), and as the recharge phase progresses, the base of the elevated convection moves lower in altitude and the virga falls lower in the atmosphere until it begins to evaporate within the radiatively-heated layer (Fig. 3a--c). The arrival of virga in the inhibition layer produces evaporative cooling rates that are approximately 20 times larger in magnitude than the antecedent radiative heating, rapidly cooling and humidifying the inhibitive cap (Fig. 3e). 

The sudden weakening of the inhibition serves as a triggering mechanism that allows a small amount of surface-based convection to penetrate into the upper troposphere for the first time in several days. Once the inhibitive cap is breached, a chain reaction ensues and the discharge phase commences: vigorous convection emanating from the near-surface layer produces strong downdrafts, which spread out along the surface as ``cold pools'' (i.e., gravity currents) and dynamically trigger additional surface-based deep convection\citep{torri2015,jeevanjee2015,Feng2015,Torri2019}. This process proceeds for a few hours, until enough convective instability has been released such that air from the near-surface layer is no longer highly buoyant in the upper troposphere. The precipitation outburst dies out, and the cycle restarts with the recharge phase.

\begin{figure*}[ht!]
\centerline{\includegraphics[width=\textwidth]{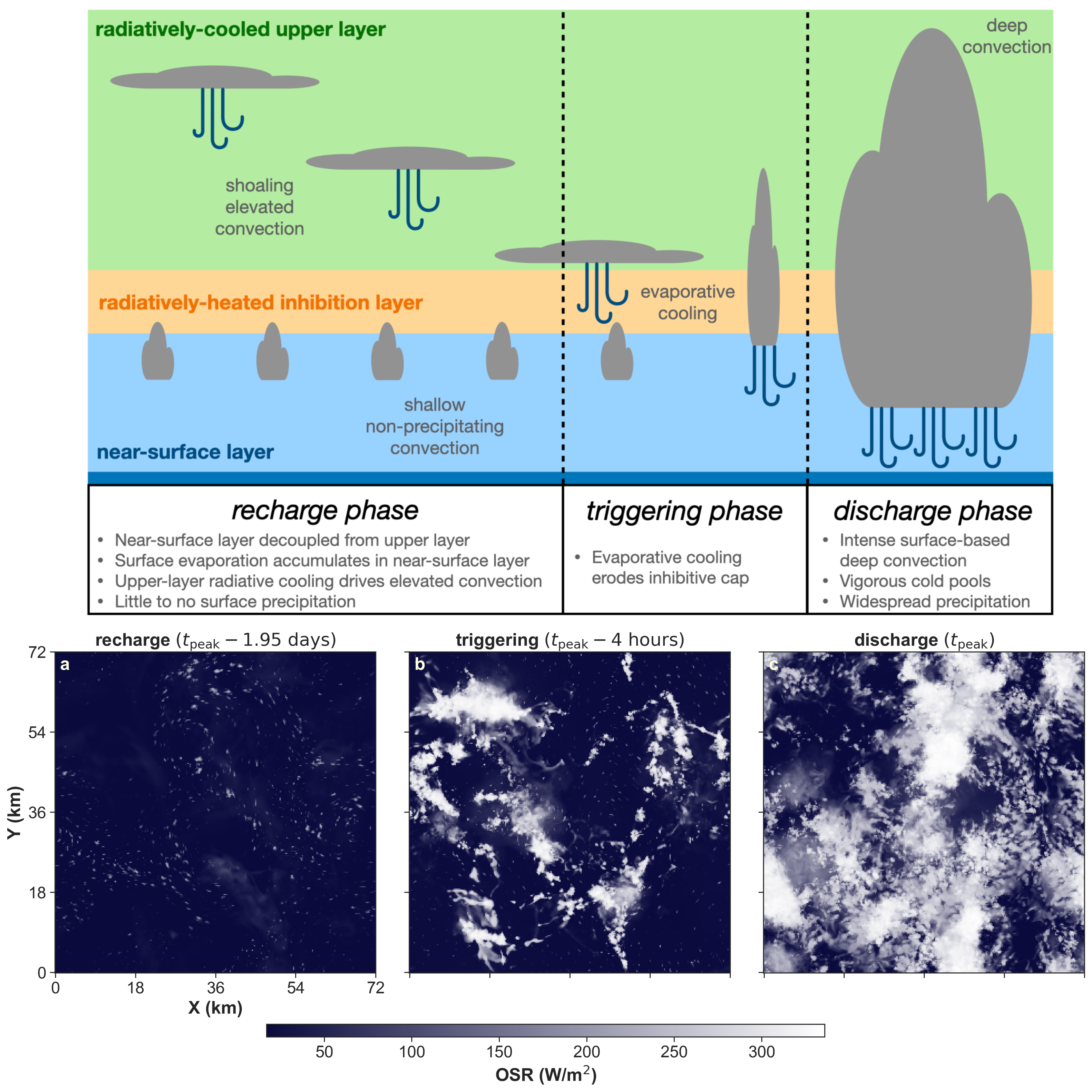}}
\caption{\textbf{Schematic view of the phases of the relaxation oscillator convective regime.} (Bottom) Snapshots of outgoing solar radiation (OSR) during the (a) recharge, (b) triggering, and (c) discharge phases, obtained 1.95 days, 4 hours, and 0 hours before the next hour of peak precipitation ($t_\mathrm{peak}$), respectively. These snapshots are from the high-resolution fixed-SST simulation at a surface temperature of 330 K. High values of OSR indicate cloud cover. Neither the graphical width of the phases nor the vertical thickness of the atmospheric layers in this schematic are proportional to the amount of time or space they occupy.}
\label{fig:schematic}
\end{figure*}

\section*{Comparison to parameterized convection}
The convectively-resolved hothouse state has both similarities and differences to prior results from models with parameterized convection. An important difference is that the time-mean temperature profile in our oscillating simulations does not resemble the three-layered structure identified in previous work\citep{Wordsworth2013,Wolf2015,Kopparapu2016,Popp2016,Wolf2018}, with a significant surface-based temperature inversion capped by a deep non-condensing layer and an overlying condensing layer further aloft. Instead, our simulations have tropospheric lapse rates that fall somewhere between the dry and moist adiabats (Fig. E4a), consistent with prior evidence that entraining moist convection sets the temperature profile in the deeply-convecting tropics\citep{Singh2013,Seeley2015}. Lacking a significant surface-based temperature inversion, our hothouse climate simulations energetically balance LTRH primarily by the latent cooling of rain evaporation rather than sensible heating of the surface. To further assess the importance of precipitation evaporation in the hothouse climate state, we modified the microphysics parameterization in the model to prevent evaporation of precipitating hydrometeors (rain, snow, and graupel). In contrast to the corresponding case with default microphysics, LTRH in the model without hydrometeor evaporation induces a mean temperature profile closely resembling the three-layered structure from previous work with parameterized convection (Fig. E4a). This suggests that the effects of evaporating hydrometeors on convective triggering and/or tropospheric energetics are critical to hothouse climates. In some global climate models (GCMs), evaporation of precipitation is either neglected or parameterized in a highly idealized manner\citep{Zhao2016}, which may be why some previous studies concluded that surface-based inversions are a defining characteristic of hothouse atmospheres\citep{Wolf2015}.

Our simulations also help clarify prior results from GCMs  regarding changes in cloud cover and climate stability in hothouse states. Previous work has suggested that LTRH causes clouds to thin or disappear from the lower troposphere and thicken in a layer of elevated convection in the upper troposphere \citep{Popp2014,Wolf2015,Wolf2018}. Similarly, in our model elevated condensation and convection during the recharge phase significantly enhance (by a factor of 3--4) the mean upper-tropospheric cloud fraction (Fig. E4b), although shallow clouds do not disappear entirely. A further similarity between our simulations and results from GCMs is the existence of a transient climate instability (i.e., a temporary sign reversal of the climate feedback parameter) during the transition to the new state induced by LTRH\citep{Popp2016,Wolf2015,Wolf2018}. In our model, the instability consists of a clear-sky longwave feedback driven by enhanced upper-tropospheric RH, which is significantly amplified by the increase in upper-tropospheric cloud cover in the oscillatory state (Fig. E5). Even for resolved convection, the net cloud radiative effect is sensitive to model details such as the horizontal resolution and microphysics scheme\citep{Wing2020,Becker2020}, so the radiative effects of clouds in the oscillatory state deserve additional study. The region of enhanced climate sensitivity associated with the transition to the hothouse state is distinct from the climate sensitivity peak found in our model at lower temperatures\citep{Romps2020}, the latter of which has been explained in terms of clear-sky feedbacks that operate in the quasi-steady convective regime\citep{Seeley2021}. 

\section*{Analogy to spontaneous synchronization}
Spatial self-aggregation of convection, in which precipitating clouds localize in the horizontal into large and persistent clusters despite spatially-uniform forcing and boundary conditions, has received considerable attention in recent years\citep{Wing2017}. The new relaxation oscillator regime revealed by our work is an analogous state of {\em temporal} convective self-aggregation: in the absence of any time-dependent forcing, deep precipitating convection becomes spontaneously synchronized (i.e., temporally localized). The oscillatory state is synchronized in the sense that subdomains separated by hundreds of kilometers exhibit boom-bust cycles of near-surface moist static energy and spikes of precipitation that are nearly in-phase (Fig. E6). The phenomenology of this synchronized atmospheric state closely resembles that of other natural systems that exhibit spontaneous synchronization\citep{Mirollo1990}, such as mechanical metronomes on a wobbly platform\citep{Pantaleone2002} and fields of flashing fireflies\citep{Buck1966}. In such systems, the key ingredient that allows for synchronization is a coupling that tends to align the phases of sub-components. In the atmosphere, there are two obvious sources of coupling between spatially-separated sub-domains: 1) gravity waves, which rapidly homogenize temperatures in the free troposphere\citep{Bretherton1989,Edman2017}, and 2) cold pools, which dynamically trigger additional deep convection in the neighborhood of prior deep convection\citep{torri2015,jeevanjee2015}. 
To investigate this analogy further, we constructed a simple two-layer model of radiative-convective equilibrium that resembles a network of noisy pulse-coupled oscillators\citep{Mirollo1990,methods}. Just as in the convection-resolving model, this two-layer model undergoes a steady-to-oscillatory transition when the amount of convective inhibition is increased (Fig. E7).

\section*{Discussion}

The hothouse convection described here bears similarities to today's climate in the Great Plains of the central United States, where elevated mixed layers transiently suppress surface-based convection until a triggering mechanism overcomes the inhibition and intense convection ensues\citep{Carlson1983,Schultz2014,Agard2017}. Our results indicate that in hothouse climates, widespread radiatively-generated convective inhibition may shift the spectrum of convective behavior away from the quasi-equilibrium regime\citep{Arakawa1974,Raymond1997} and toward an ``outburst'' regime more similar to that of the U.S. Great Plains. Since very warm climates have strongly reduced equator-pole temperature gradients\citep{Wolf2015,Popp2016}, tropical SSTs of between 330--340 K would be accompanied by moist and temperate high latitudes that might support LTRH and the convective outburst regime over a large fraction of Earth's surface. Nonetheless, an important avenue for future work is to understand how the convective outburst regime described here interacts with large-scale overturning circulations in the tropics, as well as how this regime is expressed at higher latitudes where planetary rotation and seasonal effects play an important role in atmospheric dynamics. Convection-resolving simulations on near-global domains could address these questions, and would also shed light on the prospect of convective synchronization at scales larger than we have investigated here.

It is widely recognized that most of the geological work done by precipitation (i.e., erosion, physical weathering, or sediment transport) is associated with large rain events, such that a small number of intense storms play a larger role than many small ones\citep{Melosh2011}. Although our oscillating simulations have mean precipitation rates similar to their quasi-steady counterparts, local precipitation fluxes are dramatically enhanced in the oscillatory regime. For example, in our large-domain oscillating simulation with an SST of 330 K, watershed-sized areas ($\simeq1000$ km$^2$) regularly experience 6-hour rain accumulations of several hundred millimeters, comparable to multi-day rainfall totals along the track of landfalling tropical cyclones in the United States\citep{Villarini2011}. Such large rain accumulations do not occur at all in our quasi-steady simulations (Fig. E8). If lower-tropospheric radiative heating in very warm climates leads to similar oscillatory convective behavior over land, the dramatically increased frequency of intense precipitation events would increase the fraction of rain that is converted to runoff, and presumably cause a significant acceleration of rain-induced surface alteration. Such a shift in rainfall intensity could strengthen the silicate weathering feedback well beyond the upper limit inferred from energetic constraints on the mean precipitation rate\citep{Graham2020}, and in principle might even leave an isotopic signature in the geological record\citep{Dansgaard1964,Pausata2011}. While the $\sim$320 K SST threshold for the oscillatory transition in our model is above proxy-based estimates of peak tropical SSTs during the Phanerozoic\citep{Frieling2017}, such temperatures could have been reached in earlier periods of Earth history, such as the high-CO$_2$ climates predicted in the aftermath of Neoproterozoic Snowball events \citep{Pierrehumbert2011}. 

Finally, our work has also revealed the potentially significant influence of clouds on top-of-atmosphere radiative fluxes in hothouse climates: in our oscillating simulations, elevated condensation and convection during the recharge phase enhance cloud cover, the planetary albedo, and the longwave greenhouse effect. This could be particularly relevant to tidally-locked planets orbiting M-stars for two reasons: 1) their day-sides receive permanent instellation, and 2) the M-star spectrum is shifted toward the near-infrared. Both of these factors would enhance shortwave absorption by water vapor\citep{Wordsworth2013}, and might therefore make the oscillatory convective regime more likely to occur. Indeed, in simulations with an M-star insolation spectrum, we find that the transition to the oscillatory regime occurs at a lower SST than in our standard simulations (Fig. E9). Previous work using a global model with parameterized convection found that lower-tropospheric radiative heating thinned upper-tropospheric cloud decks and led to runaway warming on tidally-locked planets\citep{Kopparapu2016}, which runs counter to the trend in high cloud amount we find in our model. These divergent model predictions highlight the importance of investigating the runaway greenhouse transition on Earth-like and tidally-locked planets using global convection-resolving models in the future.


\clearpage

\clearpage
\section*{Acknowledgements}
We are grateful to the authors of the cloud-resolving models used in this work: David Romps, Marat Khairoutdinov, and George Bryan. We also thank Anu Dudhia for sharing with us the Reference Forward Model. The authors thank Xin Wei for conducting exploratory simulations with SAM. JTS thanks Nadir Jeevanjee, Aaron Match, Nathaniel Tarshish, and Zhiming Kuang for discussions. 

\section*{Author contributions}
JTS and RWD designed the research. JTS performed the simulations, analyzed the results, and made the figures. The manuscript was written jointly by JTS and RWD. The authors declare no competing interests.

\section*{Additional information}
Correspondence and requests for materials should be directed to JTS at jacob.t.seeley@gmail.com. Reprints and permissions information is available at www.nature.com/reprints.

\section*{Data availability}
Input data files and cloud-resolving model output associated with this work is available in a Zenodo repository at the following DOI: \url{10.5281/zenodo.5117529}.

\section*{Code availability}
Source code for the stochastic two-layer model, processing cloud-resolving model output, and generating figures is available in a Zenodo repository at the following DOI: \url{10.5281/zenodo.5117529}.

\clearpage

\renewcommand{\thefigure}{E\arabic{figure}}
\renewcommand{\thetable}{M\arabic{table}}

\section*{Methods}

\subsection*{Cloud-resolving model}
We use the cloud-resolving model DAM\citep{romps2008} to simulate nonrotating radiative-convective equilibrium (RCE). RCE is an idealization of planetary atmospheres in which radiative and convective heating rates achieve time-mean balance at each altitude\citep{wing2017b}.

All DAM simulations were conducted on square, doubly periodic domains with 140 vertical levels between the surface and the free-slip, rigid lid at 60 km. Our vertical grid spacing transitions from $\Delta z=25$ m below an altitude of 650 m, to $\Delta z=500$ m between altitudes of 5.4 and 33 km, and finally to $\Delta z=1000$ m at altitudes above 38 km. Our default horizontal resolution was $\Delta x =\Delta y= 2$ km and our default horizontal domain size was $L_x=L_y=72$ km. The exceptions to this are: 1) the transient\_SO and fixedSST\_large simulations, which used larger domains of $L_x=216$ km and $L_x=512$ km, respectively; and 2) the fixedSST\_hires simulation, which used a finer horizontal resolution of $\Delta x=250$ m (Extended data table 1). For all but the fixedSST\_hires simulations, the model time step was $\Delta t=20$~s, which was sub-stepped to satisfy a CFL condition; for the fixedSST\_hires simulations, we used $\Delta t=5$ s. Overall, our model configuration is similar to the ``RCE\_small" protocol from the RCEMIP project\citep{wing2017b,Wing2020} in which DAM participated.

Surface fluxes were modeled with bulk aerodynamic formulae. Specifically, the surface latent and sensible heat fluxes (LHF and SHF) were given by 

\begin{equation}
    \mathrm{LHF}(x,y) = \rho_1(x,y) C_D\sqrt{u_1(x,y)^2 + v_1(x,y)^2 + V^2} L_v  \left[ q_\mathrm{s}^* - q_1(x,y)\right];
\end{equation}

\begin{equation}
    \mathrm{SHF}(x,y) = \rho_1(x,y) C_D\sqrt{u_1(x,y)^2 + v_1(x,y)^2 + V^2} c_p  \left[ \mathrm{SST} - T_1(x,y)\right],
\end{equation}
\noindent where $\rho_1$, $q_1$, $u_1$, $v_1$, and $T_1$ are the density, specific humidity, horizontal winds, and temperature at the first model level, $C_D=1.5\times10^{-3}$ is a drag coefficient, $V=5$ m/s is a background ``gustiness'', $L_v$ is the latent heat of condensation, $c_p$ is the specific heat capacity at constant pressure of moist air, $q_\mathrm{s}^*$ is the saturation specific humidity at the sea surface temperature and surface pressure. The surface was given a fixed, spectrally-uniform albedo of 0.07.

For simulations with a time-evolving sea surface temperature (CTRL, FSOL, and FCO2), we used a well-mixed slab ocean with horizontally-uniform temperature and heat capacity equal to that of a liquid water layer of depth 1 m. This is a standard approach\citep{Romps2020}. We used a depth of 1 m to speed the approach to equilibrium; even though this is shallower than Earth's mixed layer, we have shown that simulations with an infinite heat capacity (fixed-SST simulations) also exhibit the oscillatory regime, which shows that using a shallow ocean does not affect our main results. At each time step, the change in the slab's internal energy was equated to the sum of an applied ocean heat sink and the net surface enthalpy and radiative fluxes into the ocean. The applied ocean heat sink is necessary because limited-area simulations of the deeply convecting tropics are in a local runaway regime\citep{Romps2011a}: the absorbed shortwave radiation exceeds the outgoing longwave radiation by about 100 W/m$^2$. In the real atmosphere, this imbalance is accommodated by oceanic and atmospheric heat export, but in a limited-area cloud-resolving model coupled to a slab ocean, the imbalance must be countered by an artificial heat sink applied to the slab ocean or else runaway warming will ensue. We obtained the magnitude of the required heat sink by diagnosing the net enthalpy flux into the ocean averaged over the final 50 days of our standard simulation with a fixed sea surface temperature (fixedSST) of 305 K. This imbalance was 104.9 W/m$^2$. Our CTRL simulation (Extended data table 1), which was branched from the end of the fixedSST simulation at 305 K but with the slab ocean and a prescribed ocean heat sink of this magnitude, had a mean SST of 305 K over the ensuing 50 days of integration, confirming that the inclusion of this heat sink closed the column (ocean+atmosphere) heat budget. The solar- and CO$_2$-induced warming experiments (FSOL and FCO2) also include this same ocean heat sink.

For all simulations, domain-mean horizontal winds were nudged to zero on a timescale of 6 hours to avoid the development of stratospheric jets. To minimize artificial gravity wave reflection off the model's rigid lid, a sponge layer was also included at altitudes above 40 km in which damping was applied to all three components of the wind field.

Additional details of the DAM simulations conducted for this work are summarized in Extended data table 1. 

\begin{landscape}
\begin{table}[]
\centerline{\includegraphics[width=1.5\textwidth]{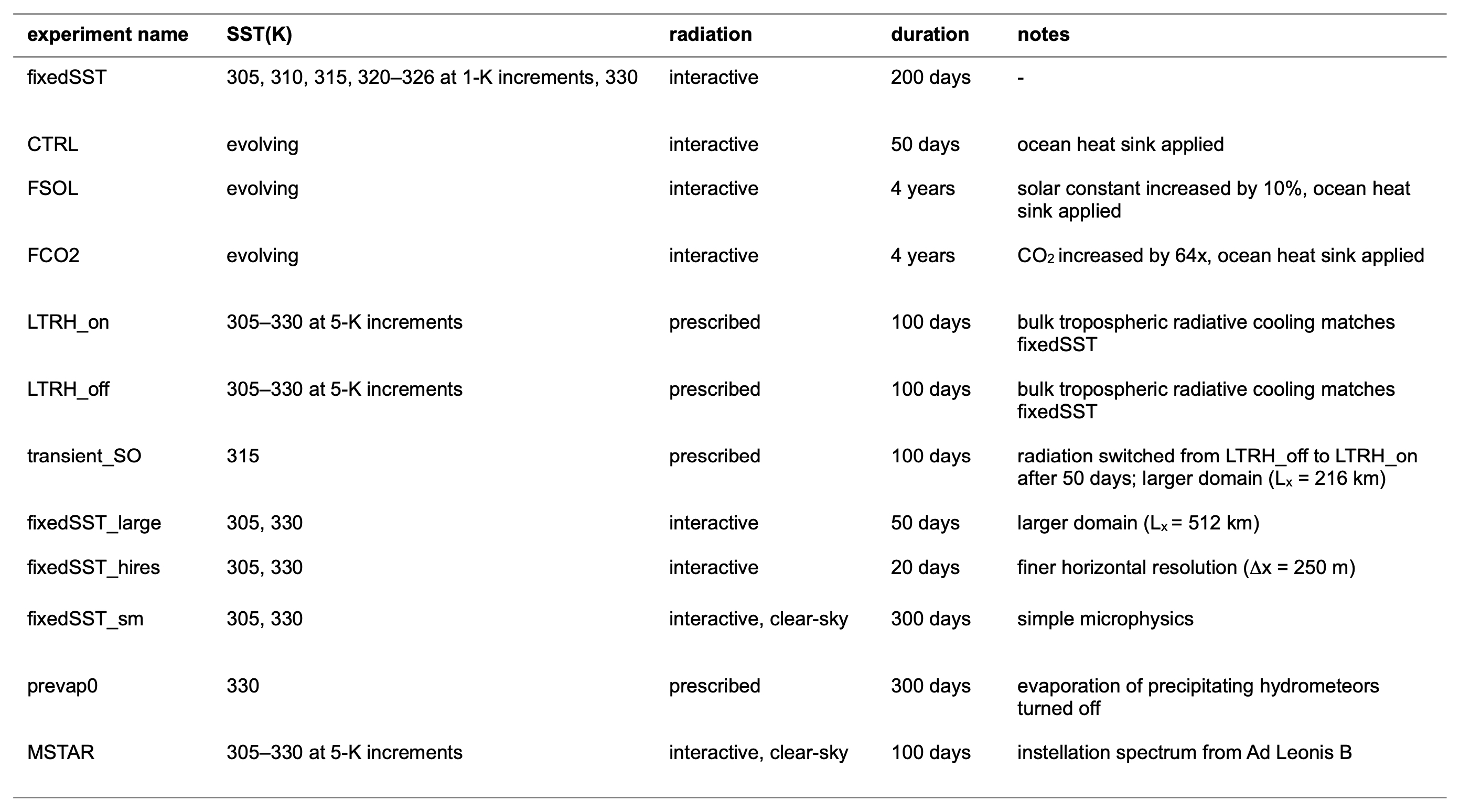}}
\caption{Summary of key aspects of the suite of DAM simulations conducted for this work.}
\end{table}\label{tab:dam_sims}
\end{landscape}

\subsection*{Radiative transfer modeling}
For shortwave and longwave radiative transfer, DAM is coupled to the fully interactive Rapid Radiative Transfer Model (RRTM)\citep{Clough2005,Iacono2008}. RRTM is a correlated-$k$ code that prioritizes computational efficiency and is validated for atmospheric conditions close to those of contemporary Earth. However, RRTM can produce unphysical results for the very warm atmospheres that are our focus. Figure E1 shows clear-sky longwave and shortwave radiative heating rates calculated by RRTM for a series of increasingly warm moist-adiabatic soundings. The unphysical discontinuities in heating rate at around 20 km altitude in the warmer soundings appear to be due to the fact that RRTM uses different lookup tables and approximations above and below a hard-coded pressure. For cooler climates, this transition pressure occurs safely in the stratosphere and there is no heating rate discontinuity, but in warmer climates the transition pressure lands in the middle of the much taller troposphere.

To obtain more realistic radiative heating rates in very warm atmospheres, we coupled DAM to a line-by-line radiation scheme known as PCM\_LBL\citep{Wordsworth2017}. This code simply solves the radiative transfer equations directly as a function of wavenumber, on a fine enough spectral grid that the heating rates in our atmospheres converge. Figure E1 shows that the clear-sky radiative heating rates calculated by PCM\_LBL closely match those of RRTM in current tropical conditions, and do not suffer from any unphysical discontinuities in heating rate in warm atmospheres. However, PCM\_LBL is a clear-sky code, so to retain the effect of cloud-radiative interactions in our simulations, we took a hybrid approach: at every call to the radiation scheme (every 200 s), we swapped out the clear-sky radiative fluxes calculated by RRTM for those calculated by PCM\_LBL for the horizontal-mean clear-sky column, while retaining the cloud-radiative effects calculated by RRTM. Using the horizontal-mean clear-sky column neglects variations in radiative heating rates due to horizontal variations in water vapor. However, sensitivity tests (not shown) indicated that this had a negligible effect on our simulation results, with differences in time-mean cloud-radiative effect of order 1 W/m$^2$. Figure E2 shows that this approach captures cloud-radiative effects closely. These effects are the dominant source of variability in top-of-atmosphere fluxes and atmospheric heating rates in our simulations.  We also note that our approach to radiation is justified by the LTRH\_on experiment suite (Extended data table 1), which shows that the transition to the oscillatory convective regime due to lower-tropospheric radiative heating does not depend on the details of cloud-radiative interactions. We validate and fully describe our radiative transfer modeling with PCM\_LBL below. 

Our longwave calculations with PCM\_LBL covered the wavenumber range from 0--4000 cm$^{-1}$, while our shortwave calculations covered 0--50000 cm$^{-1}$. The spectral resolution for both channels was 0.1 cm$^{-1}$. While this spectral resolution does not resolve the cores of lines at very low (upper-stratospheric) pressures, sensitivity tests showed that further increases in resolution yielded negligible changes to the radiative fluxes and heating rates in the troposphere, which is our focus. Similar convergence was also found in previous work \citep{Wordsworth2017,Ding2019}. At each wavenumber, the monochromatic radiative transfer equation was solved using an approach described in previous work\cite{Schaefer2016}, which uses the layer optical depth weighting scheme\cite{clough1992} to ensure accurate model behavior in strongly-absorbing portions of the spectrum. To compute radiative fluxes, we used the two-stream approximation with first-moment Gaussian quadrature \cite{clough1992}.

PCM\_LBL uses lookup tables of absorption coefficients on a pressure-temperature grid that covers the range of atmospheric conditions encountered in the model evolution, and interpolates to the current horizontal-mean atmospheric state at each vertical model level. Our pressure-temperature grid had a total of 20 pressure levels, with 10 levels spaced linearly in pressure between 110000 Pa and 10000 Pa, and 10 levels spaced logarithmically between 10000 Pa and 0.1 Pa. On each pressure level, absorption coefficients were evaluated at a set of 20 temperatures (spaced 10 K apart) that bracket the conditions encountered in the model evolution. To generate the absorption-coefficient lookup tables for H$_2$O and CO$_2$ from the HITRAN2016 database \cite{Gordon2017}, we used the RFM, a contemporary line-by-line model \cite{Dudhia2017}. Both PCM\_LBL and RRTM use MT-CKD to calculate the water vapor continuum \citep{Mlawer2012}.

For shortwave radiation, we modeled gaseous absorption only, which is appropriate for clear skies at wavelengths where Rayleigh scattering is not important. In reality, Rayleigh scattering in clear skies enhances the planetary albedo, but this process is important at significantly shorter wavelengths than the near-infrared wavelengths absorbed by H$_2$O. Therefore, the inclusion of Rayleigh scattering would introduce a small offset in the relationship between insolation and equilibrated surface temperature in our model, which would simply be absorbed into the oceanic heat sink. Our shortwave radiation setup differs for our M-star simulations (MSTAR) and for our Earth-like experiment configurations (all other simulations with interactive radiation). For our Earth-like configurations, we used top-of-atmosphere downwelling spectral solar flux data\cite{Claire2012} normalized to our specified values of the solar constant. For the MSTAR experiment, we used spectral instellation data from the M-star Ad Leonis B\citep{Segura2005}. We did not include a diurnal cycle of insolation; for our Earth-like configurations, the  cosine of the solar zenith angle was set to its insolation-weighted average during the diurnal cycle at the equator on 1 January, yielding a zenith angle of 43.75$^\circ$. With a contemporary solar constant of 1366 W/m$^2$, this yields a downwelling shortwave flux at top-of-atmosphere of 413.13 W/m$^2$. This shortwave insolation was used for all Earth-like simulations with interactive radiation except for the FSOL simulation, which used a solar constant larger by 10\%. For our MSTAR experiment, we set the cosine of the solar zenith angle to its instellation-weighted (dayside) mean, yielding a zenith angle of 48.19$^\circ$, and used a stellar constant of 800 W/m$^2$.

\subsection*{Microphysics parameterizations}\label{sec:micro}
The default microphysics scheme in DAM is known as the Lin-Lord-Krueger parameterization \citep{Krueger1995,Lin1983,Lord1984}. The LLK parameterization is a bulk scheme with six water classes (vapor, cloud liquid, cloud ice, rain, snow, and graupel). Almost all of our DAM simulations were conducted with this microphysics scheme; however, to test the robustness of our main results to microphysics, we also conducted fixed-SST simulations at 305 K and 330 K (the fixedSST\_sm suite; Extended data table 1) using a highly simplified microphysics scheme that has been described in previous work\cite{Seeley2019b, Seeley2021b}; we also describe this simplified scheme below. Additionally, the ``prevap0" simulations used the LLK microphysics parameterization, but with all evaporation of precipitating hydrometeors (rain, snow, and graupel) set to zero.

In the simplified microphysics scheme, there is no ice phase (i.e., water is modeled as a two-phase substance, with latent heat associated with phase change between vapor and liquid only). Accordingly, only three bulk classes of water substance are modeled: vapor, non-precipitating cloud liquid, and rain, with associated mass fractions $q_\mathrm{v}$, $q_\mathrm{c}$, and $q_\mathrm{r}$, respectively. Microphysical transformations between vapor and cloud condensate are handled by a saturation adjustment routine, which prevents relative humidity from exceeding 100\% (i.e., abundant cloud condensation nuclei are assumed to be present) and evaporates cloud condensate in subsaturated air. Conversion of non-precipitating cloud condensate to rain is modeled as autoconversion according to
\begin{equation}
a = -q_\mathrm{c}/\tau_\mathrm{a},
\end{equation}
\noindent where $a$ (s$^{-1}$) is the sink of cloud condensate from autoconversion and $\tau_\mathrm{a}$ (s) is an autoconversion timescale. We use $\tau_\mathrm{a}=25$ minutes, which was found in prior work to produce a similar mean cloud fraction profile as the LLK microphysics scheme\citep{Seeley2021b}. We did not set an autoconversion threshold for $q_\mathrm{c}$. Furthermore, rain is given a fixed freefall speed of 8 m/s in this simplified microphysics scheme. When rain falls through subsaturated air, it is allowed to evaporate according to

\begin{equation}
e = (q_\mathrm{v}^* - q_\mathrm{v})/\tau_\mathrm{r},
\end{equation}
\noindent where $e$ (s$^{-1}$) is the rate of rain evaporation, $q_\mathrm{v}^*$ is the saturation specific humidity, and $\tau_\mathrm{r}$ (s) is a rain-evaporation timescale. We set $\tau_\mathrm{r}=50$ hours, which was found in prior work to produce a tropospheric relative humidity profile similar to that of the LLK scheme\citep{Seeley2021b}.

\subsection*{Near-surface tracer}
In the oscillatory state, convective mass flux can be divided into two categories: updrafts that emanate from the near-surface layer ($z<1$ km), and updrafts that originate from higher in the troposphere. To discriminate between these two categories, we employed a passive tracer that measures what fraction of dry air in an updraft was recently advected from below a certain height (here, 1 km)\citep{Romps2010} and. We denote this tracer's mixing ratio as $\chi_\mathrm{ns}$. At every model time step, $\chi_\mathrm{ns}$ was set to 1 at altitudes below 1 km, and set to zero above that height except in the ``vicinity" of cloudy updrafts. We define cloudy updrafts as grid cells that have vertical velocity $w\geq0.5$ m/s and non-precipitating cloud condensate $q_n\geq10^{-2}$ g/kg, and their ``vicinity" as a cube of side length 7 grid cells centered on the updraft\citep{Romps2010}. Cloudy-updraft grid cells in which this tracer has a value of 0 contain no air that originated in the near-surface layer, so we assigned mass flux with a mean value of $\chi_\mathrm{ns}=0$ during our 5-minute sampling interval to the ``elevated" category, and assigned the remainder to the ``surface-based" category. This is an overly stringent definition of elevated convection, since any amount of surface-based convection that occurs at a certain altitude within the 5-minute sampling interval will knock other (potentially elevated) updrafts out of the ``elevated" category. Nevertheless, we still identify a large amount of elevated convection in Figure 3a of the main text.

\subsection*{Simulations with other cloud-resolving models}
In addition to our simulations with DAM, we also conducted simulations with two other cloud-resolving models: the System for Atmospheric Modeling (SAM)\citep{Khairoutdinov2003} and the Cloud Model 1 (CM1)\citep{BRYAN2002}. With each model, we conducted fixed-SST simulations at 305 K and 325 K, initialized with soundings from the corresponding fixedSST DAM simulation. 

For our SAM simulations, we used a domain of horizontal dimension 144 km with 2 km resolution. We used the same vertical grid as used for our DAM simulations, but extended to 64 km (for a total of 144 levels in the vertical) to satisfy parallelization requirements in SAM. We used a time-step of 10 s. The radiation scheme was RRTM, and the microphysics scheme was SAM's 1-moment scheme. We ran each simulation for 100 days.

For our CM1 simulations, we used a domain of horizontal dimension 72 km with 2 km resolution, and 100 vertical levels with a stretched grid (50 m resolution at altitudes below 650 m, and 500 m resolution at altitudes above 5600 m). We used a time-step of 20 s. The radiation scheme was RRTMG, and we used the Morrison double-moment microphysics scheme\citep{Morrison2005}. We ran each simulation for 150 days.

\subsection*{Stochastic two-layer model}

To explore the analogy between the oscillatory convective regime and the phenomenon of spontaneous synchronization\citep{Mirollo1990}, we constructed a simple two-layer model of RCE that resembles a network of noisy pulse-coupled oscillators. In this model, the two layers represent the near-surface layer and the upper troposphere. Each layer has a thermodynamic state variable $T$ whose time-evolution is governed by a combination of surface fluxes and convection (in the case of the lower layer) or radiation and convection (in the case of the upper layer). The upper layer is assumed to be well-mixed (with a single $T$), whereas the lower layer is divided into $N = nx \times ny$ cells arranged in a 2-dimensional, doubly-periodic lattice, each with its own $T$. Each lower-layer cell is coupled to the surface by a relaxation to a surface thermodynamic state of $T_\mathrm{s}=0$. (The choice of $T_\mathrm{s}$ is arbitrary as it simply adds an offset to the temperatures of the other layers). The two layers are also coupled by convection, which we model as a stochastically-triggered relaxation process that we describe in more detail below. 

The governing equation for the upper layer temperature $T_u$ is:

\begin{equation}
\frac{d T_u}{d t} = Q + \frac{1}{N}\sum_{i,j}M_{ij}, \label{eq:Tu}
\end{equation}

\noindent where $Q$ (K/day) is the radiative heating rate in the upper troposphere (negative values indicate cooling) and $M_{ij}$ (K/day) is the deep-convective heating rate from boundary layer cell $(i,j)$. The governing equation for the lower layer cell temperatures $T_{ij}$ is

\begin{equation}
\frac{d T_{ij}}{d t} = -T_{ij}/\tau_\mathrm{s} - M_{ij}, \label{eq:Tb}
\end{equation}

\noindent where $\tau_\mathrm{s}$ is the surface-flux timescale. Note that, due to our choice of $T_\mathrm{s}=0$, $T_{ij}$ is negative (i.e., the surface-flux term acts as a relaxation to the surface temperature of 0).

We model convective triggering and heating as stochastic processes. The generation of a convective event proceeds in three steps: 1) testing if an event is triggered; 2) determining the magnitude of the event; and 3) determining if the event can overcome an externally-specified convective inhibition parameter. For the first step, convective triggering in each grid cell is assumed to behave as a Poisson process, which is a generic representation of events that are rare and independent but that occur at an expected rate. Accordingly, the probability of convective triggering in a small time step $\Delta t$ is given by $\exp(-\lambda \Delta t)$, where $\lambda$ (day$^{-1}$) is the expected rate of convective triggering. If convection is triggered, the second step determines the size of the event by drawing an inverse timescale $\alpha_{ij}$ (day$^{-1}$) from an exponential distribution:

\begin{equation}
    \mathrm{P}(\alpha_{ij}) = \frac{1}{\beta_{ij}} \exp(-\alpha_{ij}/\beta_{ij}),
\end{equation}
\noindent for scale parameter $\beta_{ij}$. Hence larger inverse timescales, which correspond to larger convective mass fluxes/heating rates, are less common than smaller events. To represent the fact that downdrafts from neighboring convection generate larger convective plumes with lower bulk entrainment rates and larger heating rates\citep{Feng2015}, we made the scale parameter $\beta_{ij}$ linearly dependent on the convective heating rate in nearby grid cells:

\begin{equation}
    \beta_{ij} = \beta_0\left(1 + \frac{M_\mathrm{neighbors}}{M_0}\right),
\end{equation}
\noindent where $M_\mathrm{neighbors}$ is the convective heating rate summed over the ``neighborhood'' of the lower-layer cell, which we specify below. The parameter $M_0$ (K/day), as well as the chosen size of the neighborhood of each lower-layer cell, together set the sensitivity of $\beta_{ij}$ to neighboring convection. 

Finally, once the size of the triggered event is determined, $\alpha_{ij}$ is compared to the inhibition parameter $I$ (day$^{-1}$). $I$ functions as a cutoff scale: the triggered event only occurs if $\alpha_{ij}>I$. In the case of a triggered event at time $t^*_{ij}$ that overcomes the inhibition, the convective heating rate in cell $(i,j)$ is set as

\begin{equation}
    M_{ij}(t) = \begin{cases}
\max\left[\alpha_{ij}(T_{ij}-T_u),0\right], &\text{for}\quad t^*_{ij} \leq t < t^*_{ij} + \tau_\mathrm{c} \\
0, &\text{otherwise},
\end{cases} \label{eq:M}
\end{equation}

\noindent where $\tau_\mathrm{c}$ is the duration of convective events. We do not allow additional convective events to trigger when there is an ongoing event. While heuristic, this model captures some of the basic features of the convective feedbacks that occur in the full RCE simulations.

The two-layer model equations are integrated numerically with a simple forward-difference Euler method. We used parameter values $Q=-1$ K/day, $\tau_\mathrm{s}=1$ day, $\tau_\mathrm{c}=1$ hour, $\lambda=12$ day$^{-1}$, $\beta_0=1$ day$^{-1}$ and $M_0=1$ K/day. We defined the neighborhood of each grid cell as a square of side length 9 grid cells centered on cell $(i,j)$, and we used a grid of size $nx=ny=50$ and a time step of 10 minutes. We checked model convergence by halving the time step twice (to 5 and 2.5 minutes) and found very similar results in both cases.

For Figure E7, we first integrated the model for 15 days with a low value of inhibition $I=1$ day$^{-1}$, then integrated the model for two days while linearly increasing $I$ to 6 day$^{-1}$ (representing the buildup of inhibition after radiative heating is switched on in the transient\_SO simulation), and finally integrated the model for another 15 days with $I$ held fixed at the larger value.

\clearpage

\setcounter{figure}{0}  

\begin{figure*}[ht]
\centerline{\includegraphics[width=\textwidth]{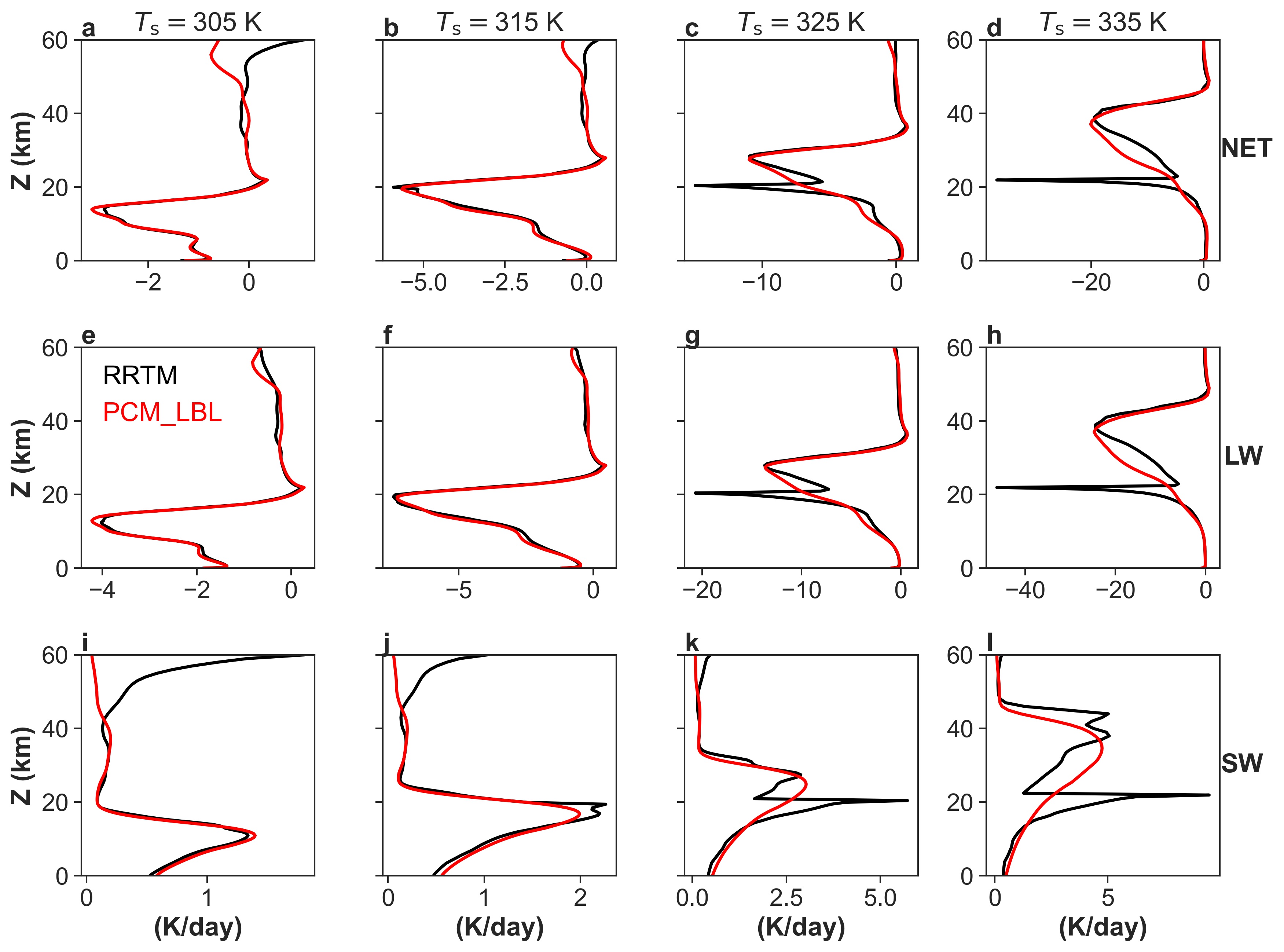}}
\caption{\textbf{Errors in clear-sky RRTM radiative heating rates are corrected by using line-by-line radiative transfer.} Comparison of net (LW+SW; panels a--d), longwave (LW; panels e--h), and shortwave (SW; panels i--l)  radiative heating rates as computed by RRTM (black) and PCM\_LBL (red). The heating rates are computed for moist-adiabatic temperature-pressure profiles with surface temperatures ranging from 305 K to 335 K in 10-K increments (columns, left to right). All columns have a surface pressure of 101325 Pa, 75\% tropospheric relative humidity, 400 ppm CO$_2$, and an isothermal stratosphere at 160 K. Note that the discontinuous heating rates calculated by RRTM for the warmer atmospheres (around 20 km altitude) do not appear in the PCM\_LBL results.}
\label{fig:benchmark_clearsky}
\end{figure*}

\begin{figure*}[ht]
\centerline{\includegraphics[width=\textwidth]{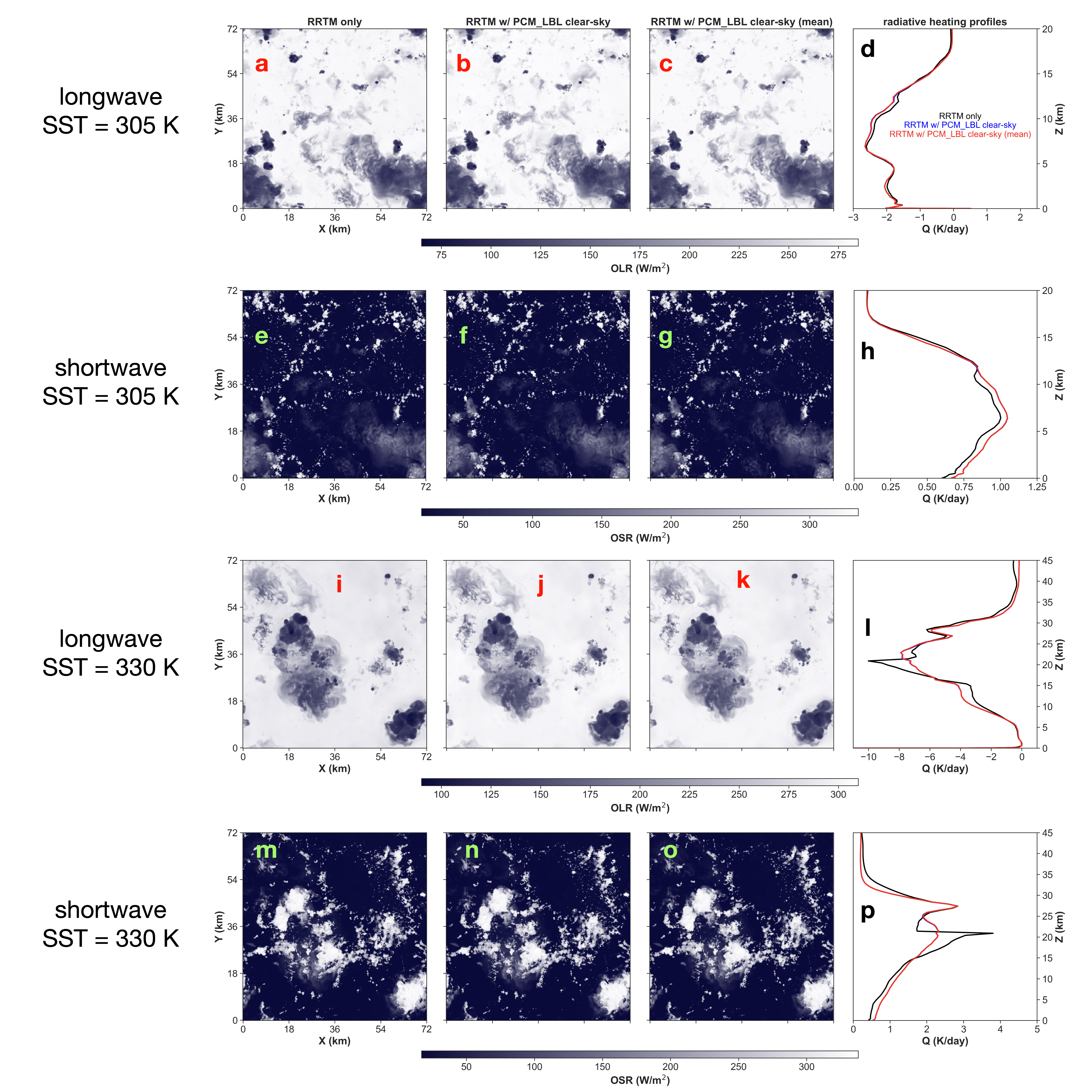}}
\caption{\textbf{Top-of-atmosphere radiative fluxes and heating rates from DAM snapshots.} (a--c) Outgoing longwave radiation (OLR) from a snapshot from the fixedSST\_hires DAM simulation with a surface temperature of 305 K, computed by three different combinations of radiative transfer codes and approximations. Panel (a) is from RRTM alone, panel (b) shows the result of swapping out the clear-sky radiative fluxes from RRTM with those calculated by PCM\_LBL, and panel (c) shows the result of swapping out each column's clear-sky radiative fluxes for those calculated by PCM\_LBL for the horizontal-mean column, which is the approach taken for the simulations associated with this work. Panel (d) shows the horizontal-mean longwave radiative heating rates for this snapshot. (e--h) As in (a--d), but for absorbed shortwave radiation (ASR). (i--p) As in (a--h), but for a snapshot from the simulation with a surface temperature of 330 K.}
\label{fig:benchmark_16}
\end{figure*}

\begin{figure*}[ht]
\centerline{\includegraphics[width=0.75\textwidth]{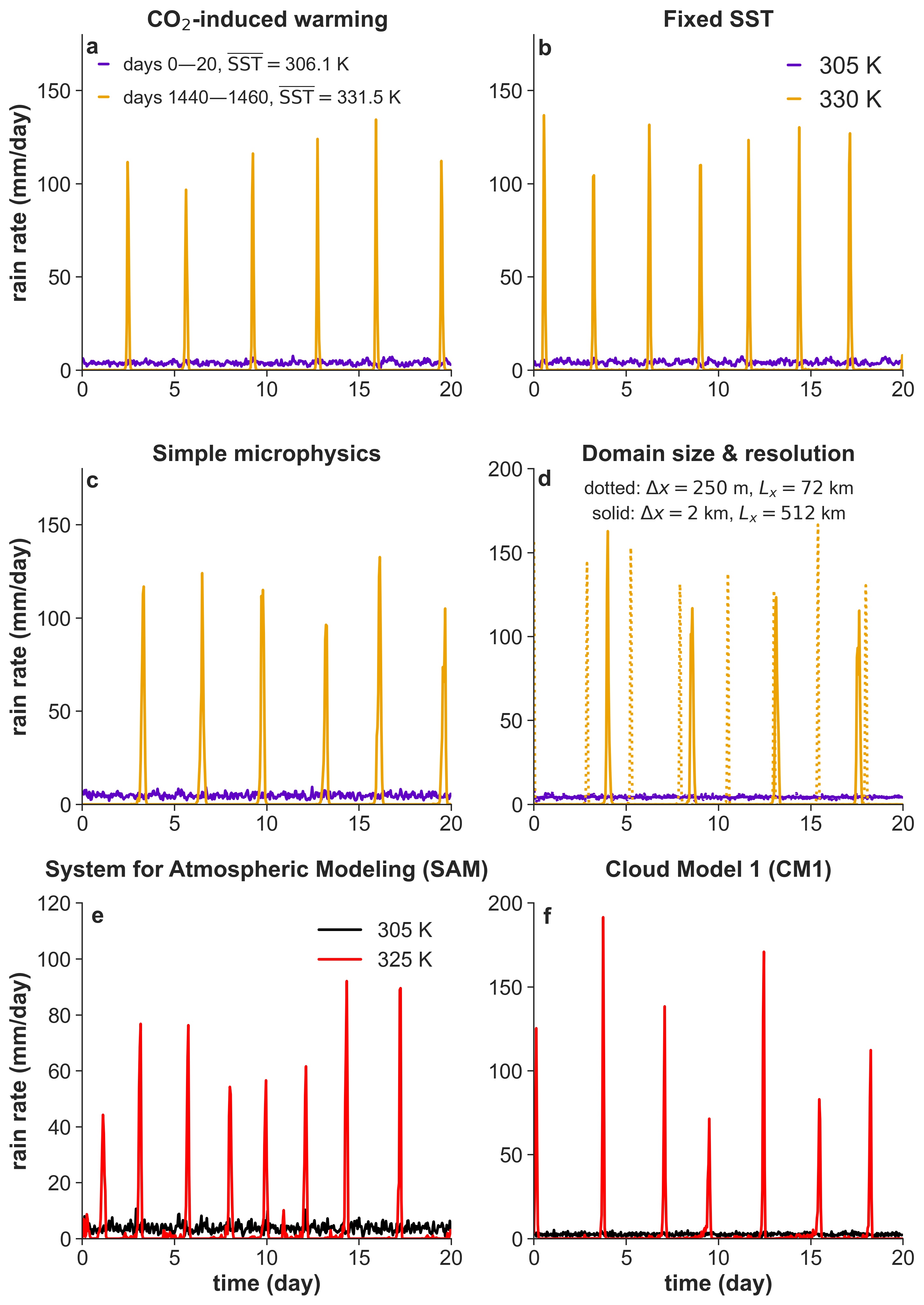}}
\caption{\textbf{Tests of the robustness of the oscillatory transition.} Domain-mean precipitation from two periods of (a) the FCO2 simulation with mean SSTs of 306.1 K and 331.5 K; (b) the fixedSST suite at 305 K and 330 K; (c) the fixedSST\_sm suite, which use the simplified microphysics parameterization described in the Methods\citep{methods}; (d) fixed-SST simulations with finer horizontal resolution ($\Delta x=250$ m; fixedSST\_hires) or on a larger domain ($L_x = 512$ km; fixedSST\_large). (e) The same quantity from  simulations conducted with the System for Atmospheric Modeling (SAM)\citep{Khairoutdinov2003} at fixed SSTs of 305 and 325 K. (f) As in (e), but for the Cloud Model 1 (CM1)\citep{BRYAN2002}.}
\label{fig:robust}
\end{figure*}

\begin{figure*}[ht]
\centerline{\includegraphics[width=\textwidth]{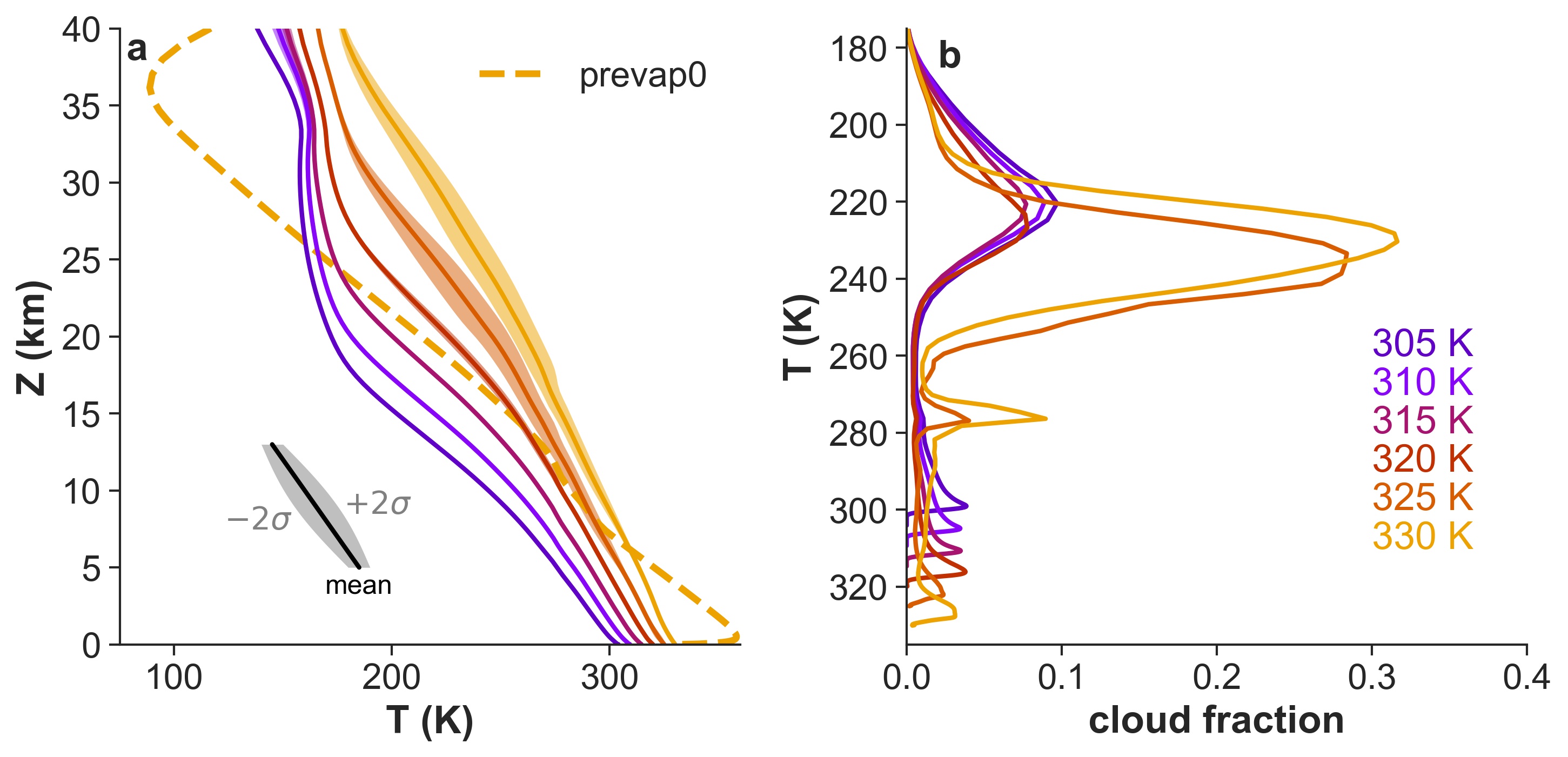}}
\caption{\textbf{Mean profiles of temperature and cloud fraction.} From the fixedSST simulations, profiles of (a) mean temperature and (b) mean cloud fraction (fraction of grid cells with non-precipitating cloud condensate mass fraction greater than 10$^{-5}$ kg/kg). In (a), the variability is indicated by the shading, which shows $\pm$2 standard deviations of hourly-mean temperatures at each altitude. In (a), the dashed line shows the mean temperature profile from the simulation without evaporation of precipitating hydrometeors (prevap0) at 330 K.}
\label{fig:tabs_cloud}
\end{figure*}

\begin{figure*}[ht]
\centerline{\includegraphics[width=0.9\textwidth]{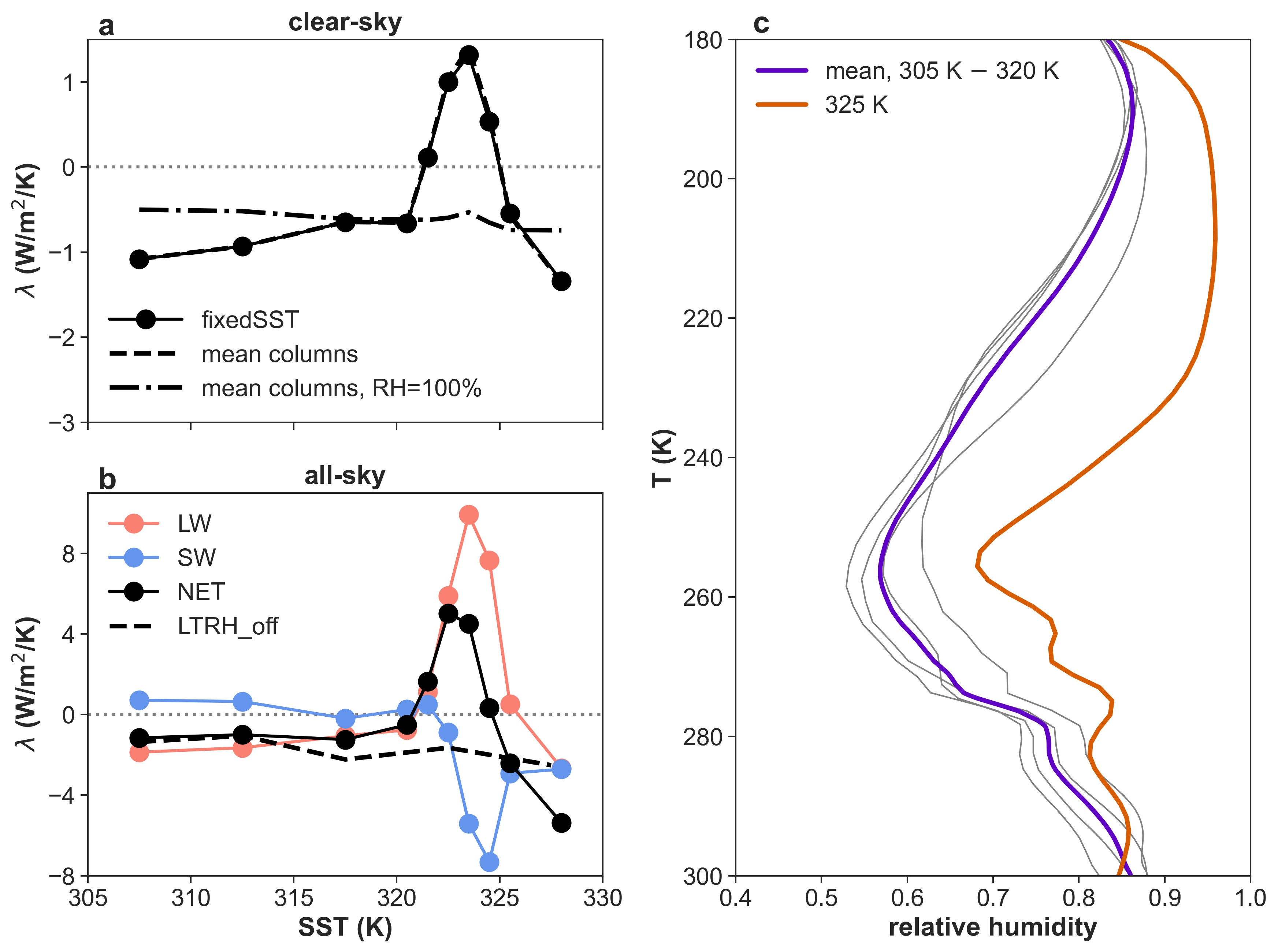}}
\caption{\textbf{Sign reversal of the climate feedback parameter indicates transient climate instability.} The feedback parameter $\lambda$ is defined here as minus the change in net radiative flux at the top-of-atmosphere (TOA) per degree of surface warming (positive downward, so that a negative feedback indicates more radiation escaping to space with warming and hence climate stability, and a positive feedback indicates climate instability; this is often called the ``Cess sensitivity"\citep{Cess1990}). We calculated feedbacks using finite differences on a staggered surface temperature grid that interpolates between the surface temperatures of the fixedSST experiment. (a) The solid line shows clear-sky feedbacks calculated for TOA fluxes averaged over the final 100 days of the fixedSST simulations, while the dashed and dot-dashed lines show the feedbacks calculated using the time-mean columns from those simulations with actual or fixed 100\% relative humidity profiles, respectively. (b) As in (a), but for the all-sky feedbacks from fixedSST experiments broken down into longwave and shortwave components. The dashed line shows the net all-sky feedback from the final 50 days of the LTRH\_off experiment, which does not undergo a steady-to-oscillatory transition and remains stable at all temperatures. (c) Time-mean profiles of relative humidity (RH) in the fixedSST experiments, using temperature within the atmosphere as a vertical coordinate to emphasize the increases in upper-tropospheric relative humidity that occur during the oscillatory transition between 320 and 325 and K. Since the clear-sky climate instability is eliminated by using a fixed relative humidity of 100\% (panel a), we attribute the clear-sky climate instability to the increase in upper-tropospheric RH, which lowers spectral emission temperatures and hence OLR.}
\label{fig:feedbacks}
\end{figure*}

\begin{figure*}[ht]
\centerline{\includegraphics[width=\textwidth]{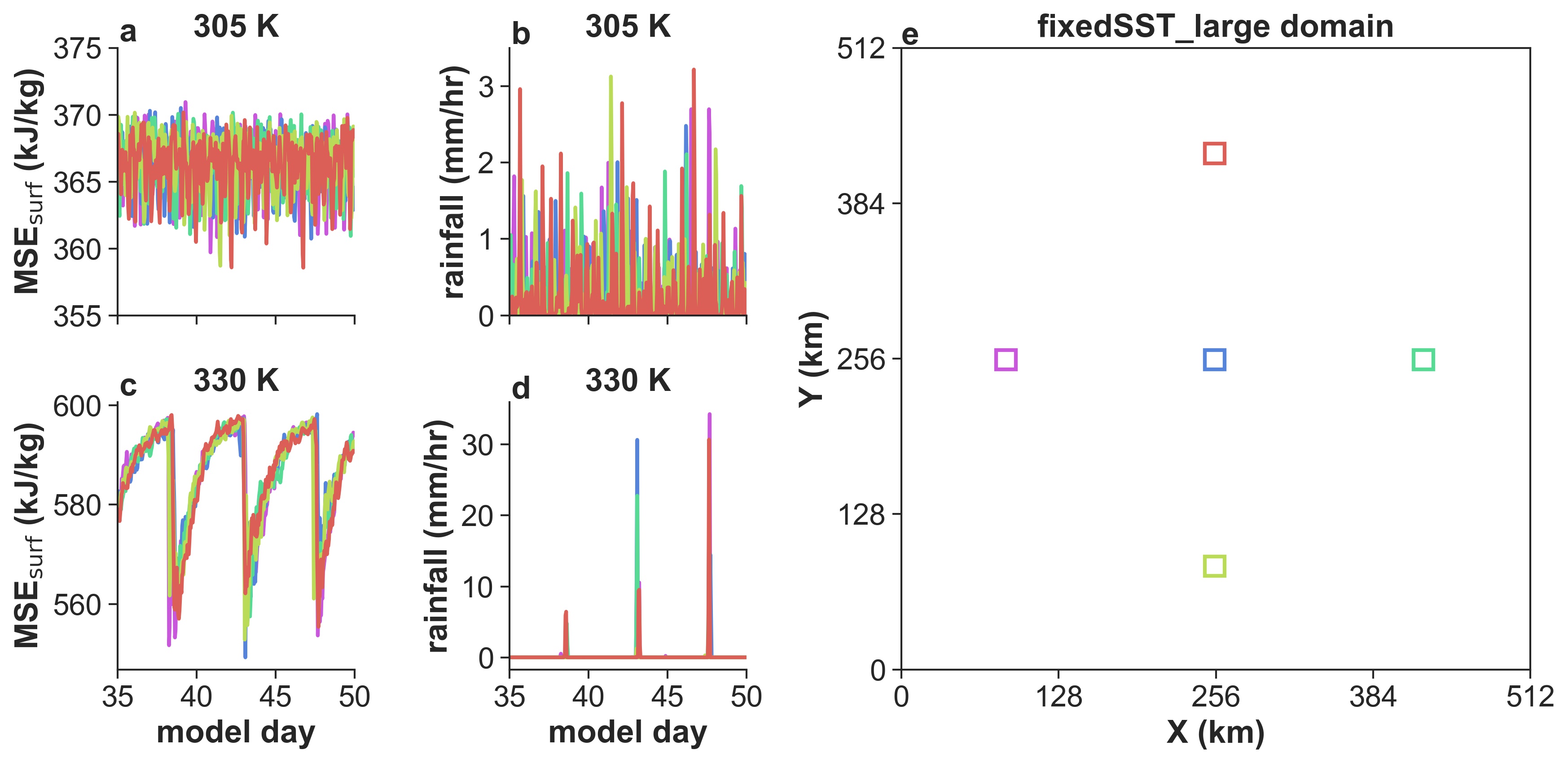}}
\caption{\textbf{Spatially-separated subdomains exhibit in-phase pulses of convection.} Timeseries of (a,c) moist static energy in the lowest model level ($z=12.5$ m; MSE$_\mathrm{surf}$), and (b,d) precipitation rate, averaged over five different subdomains of the fixedSST\_large simulations at 305 K (top row) and 330 K (bottom row). The subdomains (color-coded in panel e) each have an area of 256 km$^2$ and are located an average of 215 km apart from each other.}
\label{fig:sync_sub}
\end{figure*}

\begin{figure*}[ht]
\centerline{\includegraphics[width=0.9\textwidth]{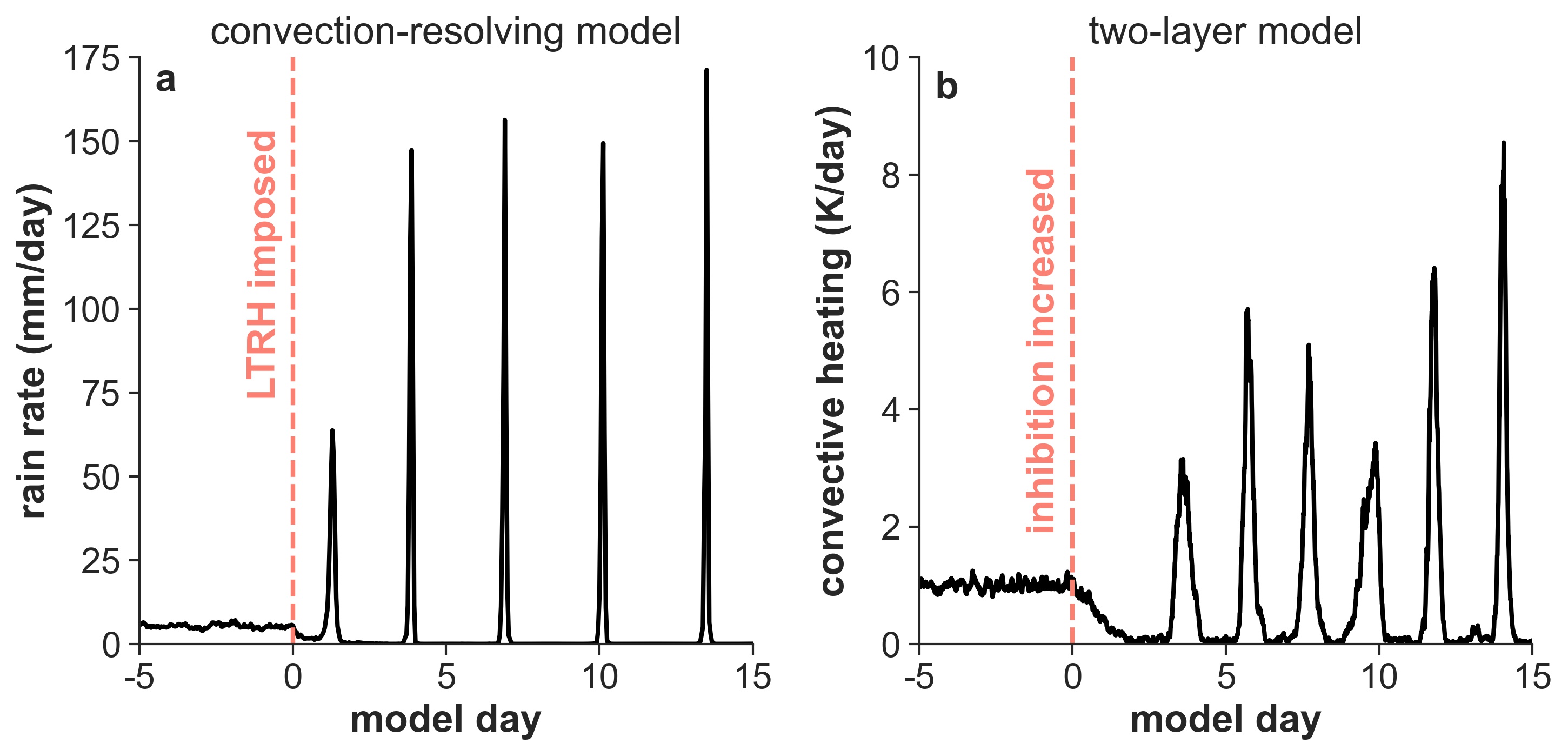}}
\caption{\textbf{The steady-to-oscillatory transition in the convection-resolving model and the stochastic two-layer model.} (a) In the convection-resolving model, the radiative heating profile is switched from cool-climate-type to hothouse-type (LTRH\_off to LTRH\_on) on model day 0 (the transient\_SO simulation). (b) In the two-layer model, the inhibition parameter is increased linearly in time between days 0 and 2 and held fixed thereafter.}
\label{fig:twolayer}
\end{figure*}

\begin{figure*}[ht]
\centerline{\includegraphics[width=\textwidth]{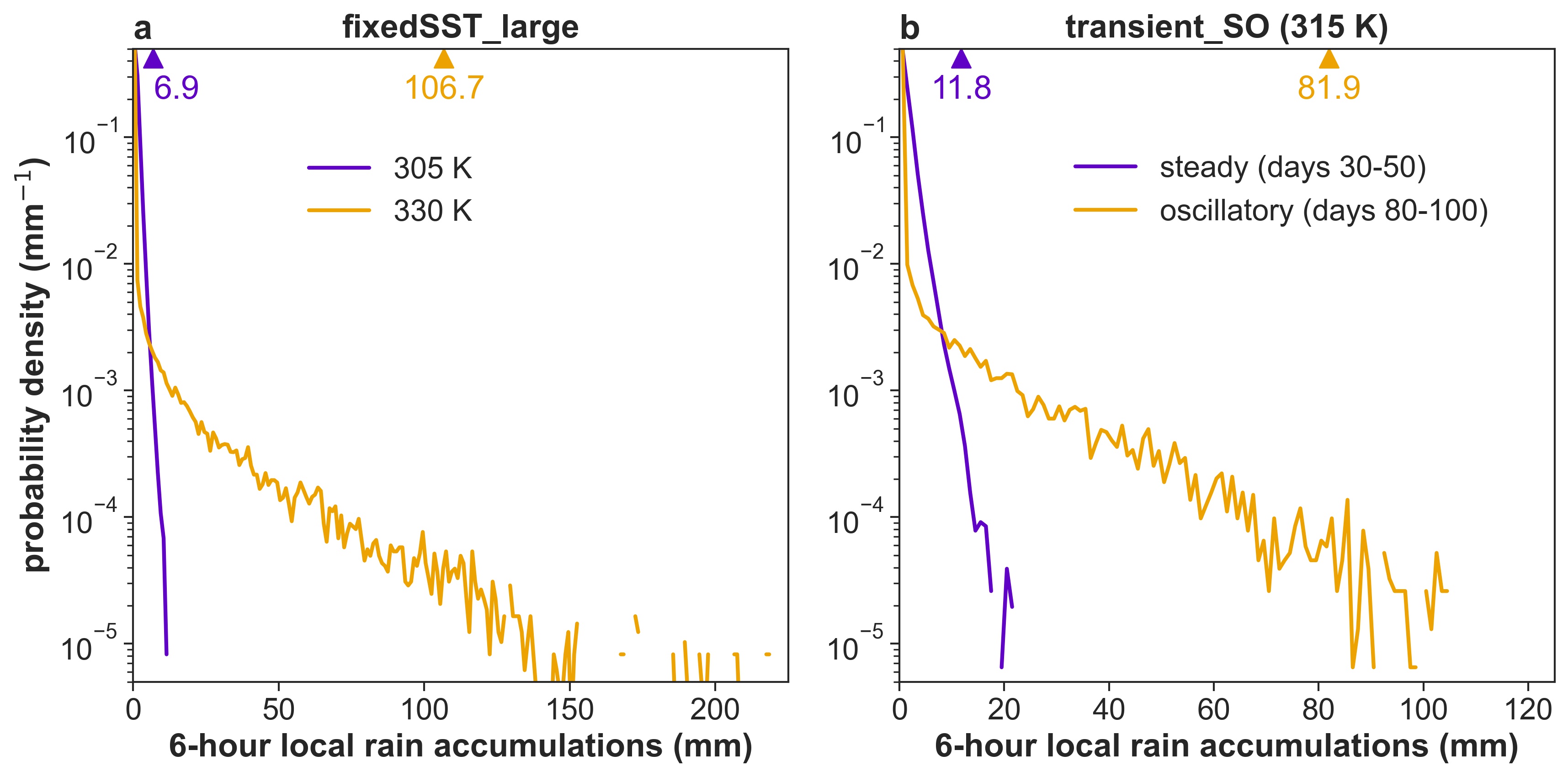}}
\caption{\textbf{Probability density functions (PDFs) of 6-hour local rain accumulations.} The precipitation data are from 20-day periods of (a) the fixedSST\_large simulations, and (b) the transient\_SO simulation in the steady and oscillatory regime. The PDFs are constructed by first dividing the model domains into watershed-sized subdomains (16$\times$16 km$^2$ for fixedSST\_large, and 12$\times$12 km$^2$ for transient\_SO). Precipitation is then accumulated in each subdomain for all 6-hour periods during the 20-day intervals, producing the 6-hour local rain accumulations from which the PDFs are constructed. The 99.9th percentile of each of the PDFs is indicated at the top of each plot.}
\label{fig:watersheds}
\end{figure*}

\begin{figure*}[ht]
\centerline{\includegraphics[width=0.75\textwidth]{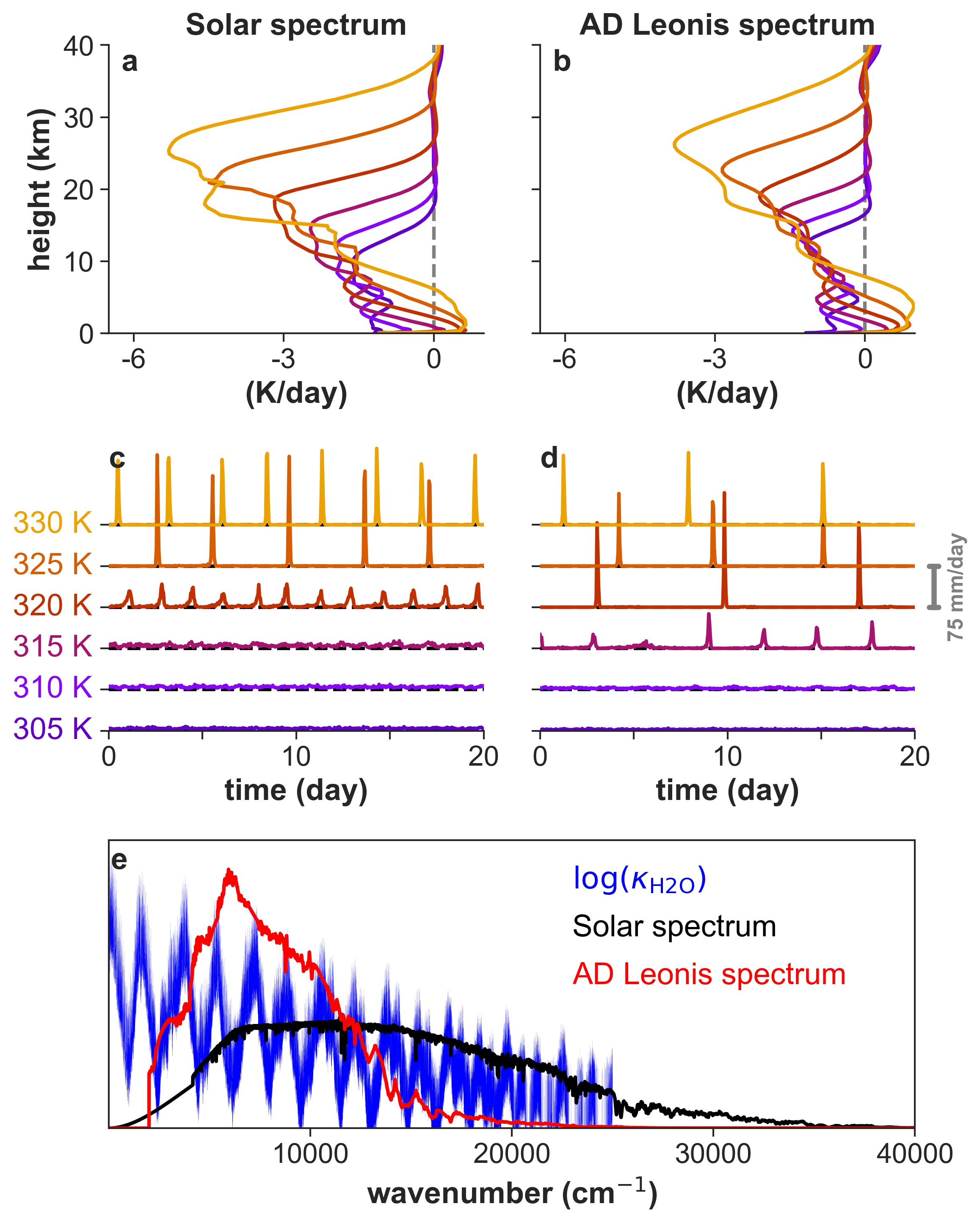}}
\caption{\textbf{The oscillatory transition occurs more readily for climates instellated by an M-star spectrum.} Comparison of tropospheric radiative heating rates (panels a,b) and timeseries of surface precipitation (panels c,d) in fixed-SST simulations with either the solar instellation spectrum or that of the M-star AD Leonis\citep{Segura2005}. Panel (e) shows the spectral flux for these two stars (normalized to the same total flux), as well as the logarithm of the H$_2$O absorption coefficient at a reference temperature and pressure.}
\label{fig:Mstar}
\end{figure*}

\clearpage
\section*{Extended data video}

An extended data video is available at the following URL:
\url{https://youtu.be/NALhYFiaeos}.

Extended data video 1 is an animation of DAM model output from the fixed-SST simulation at 330 K (from our fixedSST\_hires suite; Extended data table 1). Each frame in the video consists of 6 panels showing, from top to bottom and left to right: buoyancy in the near-surface layer, wind speed in the near-surface layer, outgoing solar radiation, temperature anomaly in the near-surface layer, specific humidity anomaly in the near-surface layer, and accumulated rainfall over the preceding 6 hours. Anomalies are calculated with respect to the horizontal- and time-mean. The sampling interval between frames is 15 minutes, and the animations cover 7 days of model time.

\clearpage

\makeatletter
\apptocmd{\thebibliography}{\global\c@NAT@ctr 50\relax}{}{}
\makeatother


\end{document}